# Electric polarizability in the three dimensional problem and the solution of an inhomogeneous differential equation

M. A. Maize[a)] and J. J. Smetanka[b)]

Department of Physics, Saint Vincent College, Latrobe, PA 15650 USA

In previous publications, we illustrated the effectiveness of the method of the inhomogeneous differential equation in calculating the electric polarizability in the one-dimensional problem. In this paper, we extend our effort to apply the method to the three-dimensional problem. We calculate the energy shift of a quantum level using second-order perturbation theory. The energy shift is then used to calculate the electric polarizability due to the interaction between a static electric field and a charged particle moving under the influence of a spherical delta potential. No explicit use of the continuum states is necessary to derive our results.

## I. INTRODUCTION

In previous work[1-3], we employed the simple and elegant method of the inhomogeneous differential equation[4] to calculate the energy shift of a quantum level in second-order perturbation theory. The energy shift was a result of an interaction between an applied static electric field and a charged particle moving under the influence of a one-dimensional bound potential. The electric polarizability, $\alpha$, was obtained using the basic relationship $\Delta E_0 = \frac{1}{2}\alpha\,\epsilon^2$ with $\Delta E_0$ being the energy shift and $\varepsilon$ is the magnitude of the applied electric field.

The method of the inhomogeneous differential equation devised by Dalgarno and Lewis[4] and discussed by Schwartz[5] can be used in a large variety of problems as a clever replacement for conventional perturbation methods. As we learn in our introductory courses in quantum mechanics, calculating the energy shift in second-order perturbation using conventional methods involves a sum (which can be infinite) or an integral that contains all possible states allowed by the transition. Some of these states, for example, continuum (scattering) states, can be fairly complicated, very difficult or impossible to obtain in a large number of problems. The knowledge of the unperturbed state will be all that we need for calculating the exact energy of that particular state to second order when we apply the techniques of the inhomogeneous differential equation.[4,5] In addition to avoiding the inclusion of the continuum in our calculations, the Dalgarno – Lewis method allowed us to study physical features which we cannot investigate with the conventional perturbation methods. Among such features was the separation of the contribution to the polarizability from the classically forbidden and classically allowed regions of the finite potential well.[3]

With our previous work and Ref. (6) as examples, a number of one-dimensional problems were treated and studied with the application of the method of the inhomogeneous differential equation. It is then reasonable to consider the three-dimensional problem. In textbooks[7-10], the three-dimensional problem appears as the computation of the electric polarizability in the case of the Hydrogen atom. The problem is certainly simple and educational. However it is the only example and the unperturbed quantum level is the ground-state of the Hydrogen atom, which is accessible and relatively easy to handle. As for the educational journals, the application of the method of the inhomogeneous differential equation to the three-dimensional problem is barely noticeable.

In this paper, we use the method of the inhomogeneous differential equation to calculate the electric polarizability of a particle moving under the influence of a spherical delta potential. This potential model has been discussed in Ref. (11) and used in Refs. (12, 13) to study photodetatchment of the negative $C_{60}^{-}$ ions. As the potential has its applications, our purpose is not to use it to do research in any area. Our first goal is to add examples demonstrating the effectiveness and simplicity of the use of the Dalgarno – Lewis method in the three-dimensional case. As we will see in this paper, the Dalgarno – Lewis technique will allow us to solve the problem analytically while avoiding approximations and heavy calculations. This in turn will enable the reader to check our work and follow the physics associated with the mathematical steps. Second is to show how the gained experience is our courses of quantum mechanics, electrodynamics, and mathematical physics can be applied to solving problems which are beneficial to many scientific areas. In the next section we will demonstrate the elegance of the Dalgarno – Lewis method in replacing the whole spectrum of eigenstates by one state, the unperturbed state, $\psi_0$, in calculating $\Delta E_0$. We will also show how to derive $\psi_0$ without dealing with the mathematical difficulties of obtaining the complete solution of the Schrödinger equation. In section III, we derive the expression for the electric polarizability and show how to check our result. In section IV, we present and discuss some of our numerical results. We close with a few concluding remarks.

## II. THE INHOMOGENEOUS DIFFERENTIAL EQUATION AND THE WAVEFUNCTION

How can the technique of the inhomogeneous differential equation be more effective than the conventional method in determining the energy eigenvalues to second-order in perturbation theory? To answer the question we begin with the conventional method. The ground state energy shift to the second-order $\Delta E_0$ is given by:

$$\Delta E_0 = \sum_{n=1}^{\infty} \frac{\langle \psi_0 | \hat{H} | \psi_n \rangle \langle \psi_n | \hat{H} | \psi_0 \rangle}{E_0 - E_n} \tag{1}$$

Here $\psi_0$ is the ground state of the unperturbed system and is occupied by a charged particle as we assume in our problem. $E_0$ is the ground state energy. The functions $\psi_n$ represent the excited states of the unperturbed system and $\grave{H}$ is the perturbation Hamiltonian. Next, we will give the basic equations for the Dalgarno – Lewis technique. Interested readers can find the details of the derivation of the inhomogeneous differential equation in Refs. (4, 5, 7, and 8).

When applying time-independent perturbation theory, the first-order correction $\phi$ to the unperturbed state $\psi_0$ can be written as:[7]

$$| \phi \rangle = \sum_{n=1}^{\infty} \frac{| \psi_n \rangle \langle \psi_n | \grave{H} | \psi_0 \rangle}{E_0 - E_n} \tag{2}$$

The ket $| \phi \rangle$ satisfies the inhomogeneous differential equation,

$$E_0 - H_0 | \phi \rangle = \grave{H} \mid \psi_0 \rangle - \langle \psi_0 | \grave{H} | \psi_0 \rangle \mid \psi_0 \rangle , \tag{3}$$

where $H_0$ is the unperturbed Hamiltonian. Operating with $\grave{H}$on Eq. (2) and then taking the inner product with $\psi_0$, we obtain the expression in Eq. (1), so that $\Delta E_0$ can be written as

$$\Delta E_0 = \langle \psi_0 | \grave{H} | \phi \rangle . \tag{4}$$

We solved Eq. (3) for $\phi$ analytically in our previous work[1-3], and an analytical solution will be obtained in this work. $\phi$ will then be used in Eq. (4) to find $\Delta E_0$. Since $\psi_0$ is the only stationary state we need, the infinite series of Eq. (1) is completely avoided and all the complications arising from dealing with scattering states disappear.

The ground state wavefunction $\psi_0$ is obtained from the three-dimensional Schrödinger equation given by:

$$\left[ \frac{-\hbar^2}{2m} \nabla^2 + V(r) \right] \psi(\vec{r}) = E \psi(\vec{r}), \tag{5}$$

with the attractive spherical-delta potential $V(r)$ is given by

$$V(r) = g \, \frac{\delta(r - r_0)}{r_0^2}. \tag{6}$$

Here g in our problem is an adjustable parameter and is always negative.

$V(r)$ represents an interaction which is zero everywhere except at the surface of a sphere of radius $r_0$. The function $\frac{\delta(r-r_0)}{r_0^2}$ has the same units as the Dirac delta $\delta(\vec{r})$ but unlike $\delta(\vec{r})$, it has eigenstates which are finite everywhere. As discussed in Ref. (14), the wavefunction produced by the non-perturbative treatment of the Schrödinger equation in the case of $\delta(\vec{r})$ will not be finite at the origin.

Since $V(r)$ is spherically symmetric, we can use the separation of variables to write $\psi(\vec{r}) = R_l(r)Y_{l,m}(\theta, \phi)$. The radial part of Eq. (5) can then be written as

$$\left[\frac{1}{r^2}\frac{d}{dr}r^2\frac{d}{dr} - \frac{\gamma}{r_0}\delta(r - r_0) - \frac{l(l+1)}{r^2}\right]R_l(r) = \left(\frac{-2mE}{\hbar^2}\right)R_l(r), \quad (7)$$

where the dimensionless quantity $\gamma = \left(\frac{2\,m\,g}{\hbar^2\,r_0}\right)$. The three-dimensional Schrödinger equation with the spherical delta potential produces a spectrum of bound states in addition to a continuum of unbound states.

Now, if we want to follow the conventional method in calculating the energy shift to second order of the unperturbed level, our first step will be solving Eq. (7) for the eigenstates and eigenvalues (both bound and scattering) which satisfy the equation. The details of solving Eq. (7) are certainly beyond the knowledge we gain from our undergraduate courses and also far from common practice of many of us. In comparison, if we use the Dalgarno – Lewis method to find the energy shift to second order of the unperturbed state, the knowledge of the unperturbed state is all we need. In our problem, the unperturbed state is the ground state, $\psi_0$, and to calculate it we apply no more than basic calculus and undergraduate quantum mechanics.

We set $l = 0$, make the substitution $R_0(r) = \frac{Q_0(r)}{r}$ and consider the case of $r \neq r_0$ all in Eq. (7), yielding

$$\frac{d^2Q_0(r)}{dr^2} = k_0^2 Q_0(r), \quad (8)$$

where we take $E$ to be equal to $E_0 = \frac{-\hbar^2 k_0^2}{2\,m}$. The solution for $Q_0(r)$ from Eq. (8) is in the form $e^{\pm k_0 r}$. For $r < r_0$, $Q_0(r < r_0) = A_1 \sinh k_0 r$, so $Q_0(r < r_0) \to 0$ $as\, r \to 0$ and $A_1$ is a constant. For $r > r_0$, $Q_0(r > r_0) = A_2 \mathrm{e}^{-k_0 r}$, so $Q_0(r > r_0) \to 0$ $as\, r \to \infty$ and $A_2$ is a constant. $A_1$ and $A_2$ are determined by applying the continuity of the wavefunction at $r = r_0$ and normalization. With $A_1$ and $A_2$ determined, the wavefunction is given by

$$Q_0(r < r_0) = N_0 e^{-k_0 r_0} \sinh k_0 r, \quad (9)$$

and

$$Q_0(r > r_0) = N_0 \sinh k_0 r_0\, \mathrm{e}^{-k_0 r}, \quad (10)$$

where

$$N_0 = 2\sqrt{\frac{k_0}{1-(1+2k_0 r_0)e^{-2k_0 r_0}}}. \quad (11)$$

To calculate $E_0$, we return to Eq. (7), set $R_l(r) = \frac{Q_l(r)}{r}$ and integrate the equation from $r_0 - \Delta$ to $r_0 + \Delta$ with $\Delta \to 0$ yeilding

$$\left[\frac{dQ_l(r > r_0)}{dr}\right)_{r = r_0+\Delta} - \frac{dQ_l(r < r_0)}{dr}\Big)_{r = r_0-\Delta}\Big]_{\Delta \to 0} = \left(\frac{\gamma}{r_0}\right) Q_l(r_0). \quad (12)$$

Eq. (12) gives the discontinuity condition of the derivatives of all functions and we get the ground state equation by setting $l = 0$. Now substituting $Q_0(r)$ from Eq. (9 and 10) in Eq. (12) produces the transcendental equation providing $E_0$ given by

$$\frac{-2\, k_0 r_0}{\gamma} = (1 - e^{-2\, k_0 r_0}), \quad (13)$$

keeping in mind that $\gamma$ is negative.

## III. THE ELECTRIC POLARIZABILITY

Since we have $|\, \psi_0 \rangle$ and $E_0$, we need to determine the perturbation Hamiltonian $\grave{H}$ to solve Eq. (3) for $|\, \phi \rangle$. The electric dipole interaction produces the perturbation Hamiltonian $\grave{H} = -q\, \varepsilon\, r\, Cos\theta$, where the electric field is applied in the z direction. When we refer to the right hand side of Eq. (3), we find that the matrix element $\langle \psi_0 | \grave{H} | \psi_0 \rangle$ goes to zero due to parity considerations and accordingly the right hand side is proportional to $Cos\theta$. We then can write $|\, \phi \rangle$ in coordinate representation as $\phi(\vec{r}) = P(r) Cos\theta$. By setting $P(r) = \frac{s(r)}{r}$, we can write Eq. (3) for the region $r < r_0$as

$$\left(-k_0^2 + \frac{d^2}{dr^2} - \frac{2}{r^2}\right) s(\, r < r_0) = -q\, \varepsilon\, r\, \acute{N} \left(\frac{2m}{\hbar^2}\right) e^{-\, k_0 r_0}\, Sinh\, k_0 r\, . \quad (14)$$

The term $\frac{2}{r^2}$ on the left hand side of Eq. (14) comes from the term $\frac{l\,(l+1)}{r^2}$ since $Cos\theta$ is proportional to the spherical harmonic with $l = 1$, $m_l = 0$ and $\acute{N} = \frac{N_0}{\sqrt{4\pi}}$. The first step in determining the solution for an inhomogeneous differential equation of the second order is to find two linearly independent solutions of the homogeneous version of the equation. It is not difficult to predict that the solutions will be a multiplication of $Cosh$ and $Sinh$ functions with functions of $\left(\frac{1}{r}\right)^n$. With some trial and error, the two linearly independent solutions are $s_a(r) = Cosh\, k_0 \mathrm{r} - \frac{Sinh\, k_0 \mathrm{r}}{k_0 \mathrm{r}}$ and $s_b(r) = Sinh\, k_0 \mathrm{r} - \frac{Cosh\, k_0 \mathrm{r}}{k_0 \mathrm{r}}$. The next step is to use the functions $s_a(r)$ and $s_b(r)$ and their Wronskian, $W[s_a(r), s_b(r)] = s_a(r) \left(\frac{ds_b(r)}{dr}\right) - s_b(r) \left(\frac{ds_a(r)}{dr}\right)$ in performing straight forward integration to complete the solution of the inhomogeneous equation.[15] After eliminating the terms which diverge when $r \to 0$, $s(\, r < r_0)$ can be written as

$$s(\,r<\,r_0)=\frac{G_1}{8k_0^4r}[(-3k_0r+2k_0^3r^3)Coshk_0r+3\,Sinh\,k_0r] \quad (15)$$
$$+\frac{C}{k_0r}(k_0r\,Cosh\,k_0r-Sinh\,k_0r),$$

where $G_1=\ -q\,\varepsilon\,\acute{N}\left(\frac{2m}{\hbar^2}\right)\,e^{-\,k_0r_0}$ and $C$ is a constant which will be determined.

Moving to the region $r>r_0$ and following the same steps which lead to Eq. (14), we can write the inhomogeneous differential equation for $s(\,r>\,r_0)$ as

$$\left(-k_0^2+\frac{d^2}{dr^2}-\frac{2}{r^2}\right)s(\,r>\,r_0)=\ -q\,\varepsilon\,r\,\acute{N}\left(\frac{2m}{\hbar^2}\right)\,Sinh\,k_0r_0\,e^{-\,k_0r}, \quad (16)$$

and after solving the differential equation and then eliminating the terms which diverge when $r\to\infty$, $s(\,r>\,r_0)$ can be written as

$$s(\,r>\,r_0)=\frac{G_2}{8k_0^4r}[(3+3k_0r-2k_0^3r^3)e^{-k_0r}]+\frac{D}{k_0r}(k_0r+1)e^{-k_0r}, \quad (17)$$

where $G_2=\ -q\,\varepsilon\,\acute{N}\left(\frac{2m}{\hbar^2}\right)\,Sinh\,k_0r_0$ and $D$ is a constant which will be determined. The second term on the right hand side of Eq. (17) is proportional to $s_a(r)-\ s_b(r)$.

Based on Eq. (2), the continuity condition for any of the unperturbed wavefunctions at $r=r_0$ and the discontinuity condition for the derivatives [Eq. (12)] will apply to $s(\,r)$. Applying the continuity and discontinuity conditions to $s(\,r<\,r_0)$ and for $s(\,r>\,r_0)$, we obtain two equations which can then be used to find C and D. Substituting C and D in Eqs. (15) and (17) respectively we then have $\phi(\vec{r})$. $\Delta E_0$ is obtained from the integration of Eq. (4) and $\alpha=\frac{-\,2\,\Delta E_0}{\varepsilon^2}$. The expressions for C, D and $\alpha$ are given in Appendix A.

It is always beneficial to check our results, and to check our results for $\alpha$ we start with the fundamental expression given in Eq. (1). This might indicate that we need to find all possible $\psi_n$'s and $E_n$'s but all we need is the expression for the bound state for $l=1$, $\psi_1$ and the corresponding energy $E_1$. Since the interaction $\grave{H}$ is a dipole interaction, and the smallest energy gap possible is $(E_1-E_0)$, then the transition between $\psi_0$ and $\psi_1$ can be responsible for most (almost all) the contribution to $\Delta E_0$ (Eq. (1)) when the attraction is strong enough.

Using the separation of variables as we did before we write $\psi_1(\vec{r})$ as $\psi_1(\vec{r})=R_1(r)Y_{1,0}(\theta)$. Now writing $R_1(r)=\ \frac{Q_1(r)}{r}$, the radial part of the Schrödinger equation for $r\neq r_0$ is

$$\left(-k_1^2+\frac{d^2}{dr^2}-\ \frac{2}{r^2}\right)Q_1(r)=0, \quad (18)$$

where $E_1=\frac{-\,\hbar^2k_1^2}{2\,m}$. Eq. (18) is the homogeneous version of Eqs. (14) or (16), which means we already have the radial dependence function of $Q_1(r<\,r_0)$ and $Q_1(r>\,r_0)$. For $r<r_0$, $Q_1(r<\,r_0)=B_1\left(Cosh\,k_1r-\ \frac{Sinh\,k_1r}{k_1r}\right)$, so $Q_1(r<\,r_0)\to 0$ as $r\to\ 0$ and $B_1$ is a constant.

For $r > r_0$, $Q_1(r > r_0) = B_2\left(1 + \frac{1}{k_1 r}\right)e^{-k_1 r}$, so $Q_1(r > r_0) \to 0$ as $r \to \infty$ and $B_2$ is a constant. $B_1$ and $B_2$ are then determined by applying the continuity of the function at $r = r_0$ and normalization. With $B_1$ and $B_2$ determined, the wave function is given by

$$Q_1(r < r_0) = N_1\left(1 + \frac{1}{k_1 r_0}\right)\left(Cosh\, k_1 r - \frac{Sinh\, k_1 r}{k_1 r}\right)e^{-k_1 r_0} \quad (19)$$

and

$$Q_1(r > r_0) = N_1\left(Cosh\, k_1 r_0 - \frac{Sinh\, k_1 r_0}{k_1 r_0}\right)\left(1 + \frac{1}{k_1 r}\right)e^{-k_1 r} \quad (20)$$

where

$$N_1 = \sqrt{\frac{4k_1^3 r_o^2}{k_1^2 r_o^2 - 3 + (2k_1^3 r_o^3 + 5k_1^2 r_o^2 + 6k_1 r_0 + 3)e^{-2k_1 r_0}}}\,. \quad (21)$$

Substituting by $Q_1(r < r_0)$ and $Q_1(r > r_0)$ in the discontinuity equation, Eq. (12), we get the transcendental equation which determines $E_1$ and it is given by

$$\frac{-k_1 r_0}{\gamma} = \frac{Q_1(r_0)}{N_1} \quad (22)$$

Applying the wavefunctions $\psi_0$ and $\psi_1$ with the corresponding eigenvalues to Eq. (1) we get the electric polarizability due to the transition $\psi_0 \to \psi_1$ and we call it $\alpha_b$. We give the expression for $\alpha_b$ in Appendix A.

## IV. Numerical Results and Discussion

In this section, we discuss some of the physical features of our system. In addition, we illustrate the usefulness of the Dalgarno-Lewis method in explaining our results. In Fig. (1), the vertical axis represents the polarizability in units of $m^3$. The units of $m^3$ is obtained by multiplying our $\alpha$ by $\frac{1}{4\,\pi\,\varepsilon_0}$ ($9 \times 10^9$ N $m^2$ $C^{-2}$). We do so because the polarizabitliy is usually given in units of length cubed as introduced in undergraduate textbooks[16]. We choose some radius $r_0$ = 3 Å and we plot the polarizability versus the variable $|\gamma|$ in figure (1).

Figure 1: Electric Polarizability

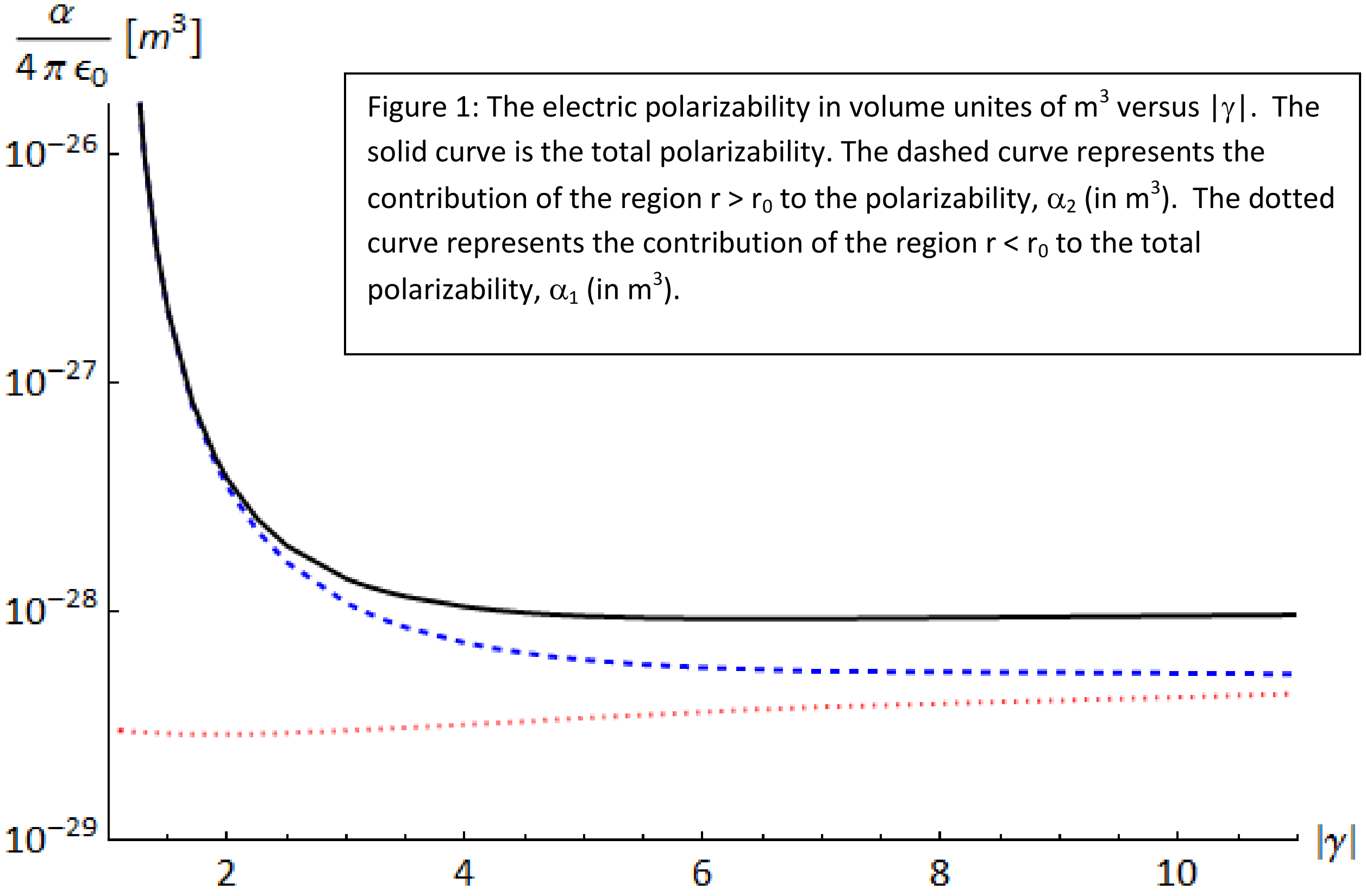

Figure 1: The electric polarizability in volume unites of $m^3$ versus $|\gamma|$. The solid curve is the total polarizability. The dashed curve represents the contribution of the region $r > r_0$ to the polarizability, $\alpha_2$ (in $m^3$). The dotted curve represents the contribution of the region $r < r_0$ to the total polarizability, $\alpha_1$ (in $m^3$).

In figure (1) the dependence of the electric polarizability on the value of $|\gamma|$ is given for $|\gamma| > 1$. There is no polarizability for $|\gamma| \leq 1$ since there are no real values for $r_0k_0$ except for $|\gamma| > 1$ as we solve Eq. (13) to obtain $r_0k_0$ for a given value of $|\gamma|$. With no real values for $r_0k_0$, there will be no real value for $k_0$ and consequently no bound states. No bound states does of course result in no polarizability. As $|\gamma|$ goes from larger but close to 1 to 3, the total electric polarizability (shown in the black solid line) drops must faster relative to the change in the total electric polarizability at large $|\gamma|$ where it decreases at a much slower rate reaching saturation. In our problem, the dipole transitions are from the ground state to the first excited state (bound state with $\ell = 1$) and from the ground state to the continuum. We do not include the continuum explicitly in our calculations. In solving Eq. (22) to get $r_0k_1$ for a given $|\gamma|$, there are no real values for $r_0k_1$ except for $|\gamma| > 3$. Accordingly, the contribution to the electric polarizability in the region of $|\gamma|$ close to but larger than 1 to $|\gamma| = 3$ is all due to the transitions between the ground state and the continuum. As $|\gamma|$ increases for fixed $r_0$, the binding potential becomes stronger leading to pushing down the energy of the ground state and increasing the energy gap between the bound particle and the continuum. The increase of the energy gap (Eq. (1)) will then lead to the decreased polarizability. As $|\gamma|$ goes beyond 3, the contribution of the bound state $\ell = 0$ to the bound state $\ell = 1$ becomes the more dominate contribution to the electric polarizability (figure (2)). The bound to bound contribution reaches saturation for large $|\gamma|$. The values for $|\gamma|$ corresponding to the appearance of bound states where given in Ref. (11) as part of the solution of Schrödinger's equation (Eq.(7)).

Figure 2: Bound to Bound Fraction of Polarizability

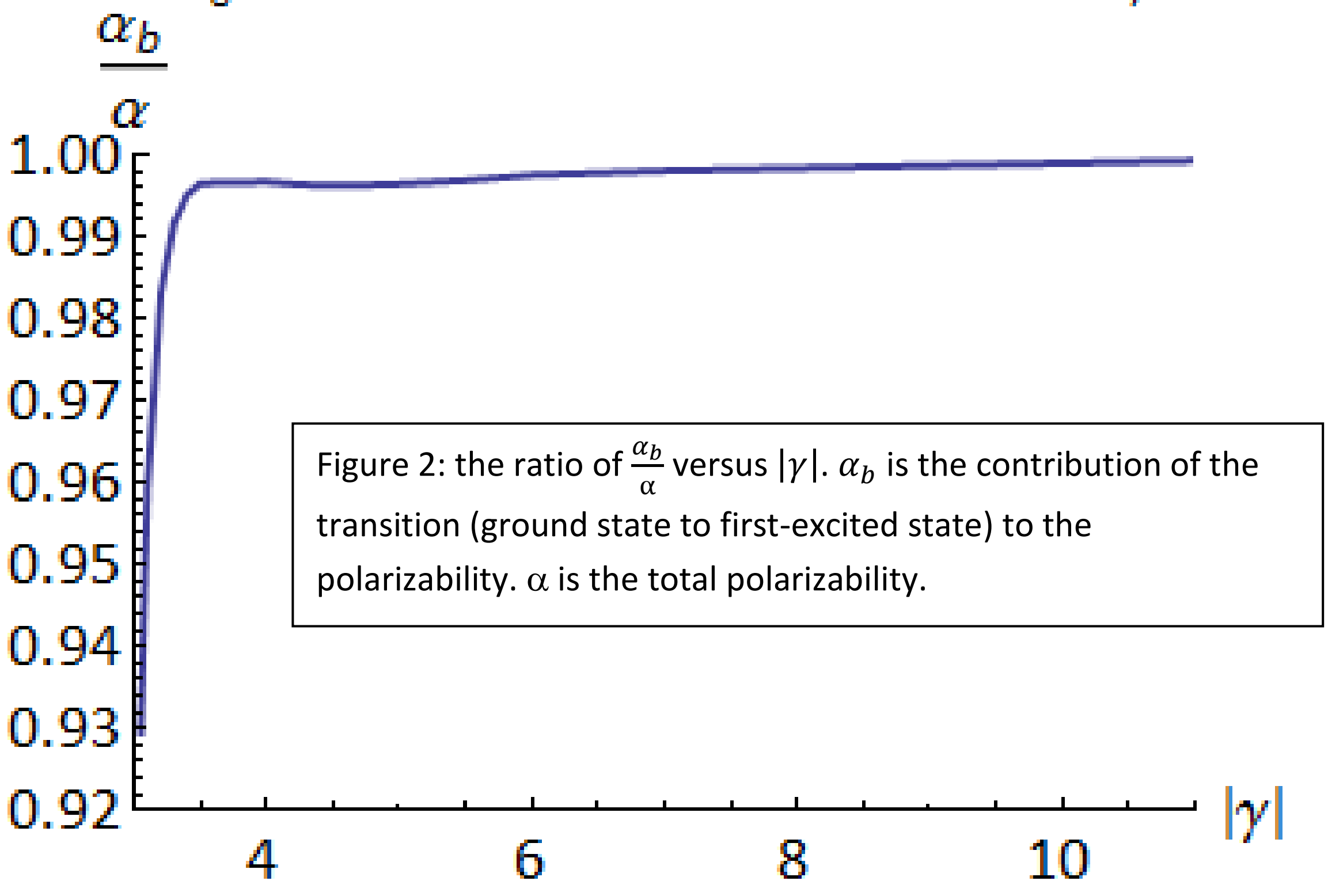

Figure 2: the ratio of $\frac{\alpha_b}{\alpha}$ versus $|\gamma|$. $\alpha_b$ is the contribution of the transition (ground state to first-excited state) to the polarizability. $\alpha$ is the total polarizability.

The use of the Dalgarno-Lewis method gives us the advantage of comparing the contributions of the system in the $r < r_0$ and the $r > r_0$ regions to the electric polarizability. Let's call the contribution from the region $r < r_0$ to the polarizability $\alpha_1$ (in $m^3$) (figure 1 dotted line) and the $r > r_0$ region contribution $\alpha_2$ (in $m^3$) (figure 1 dashed line). For $|\gamma|$ close to but larger than one to $|\gamma|$ close to three, $\alpha_2$ is much larger than $\alpha_1$ (see figure 1). With increasing $|\gamma|$, $\alpha_1$ and $\alpha_2$ approach each other with $\alpha_2$ holding a slight edge in the contribution at large $|\gamma|$ ($|\gamma| \geq 11$ in figure 1). To illustrate the reason behind the difference between $\alpha_1$ and $\alpha_2$, first we observe that the polarizability is dependent on the integration of $rQ_0(r)S(r)$ (Eq. 4). Second, we plot $Q_0(r)$ (figure (3)) and $S(r)$ (figure 4) for three different values of $|\gamma|$. Increasing $|\gamma|$ means an increase of binding which leads to $Q_0(r)$ moving to a smaller space (figure 3). In addition $Q_0(r)$ approaches symmetry around $r_0$ with increasing $|\gamma|$. Similarly, $S(r)$ also goes to smaller r and approaches symmetry around $r_0$ with increasing $|\gamma|$ (figure 4). The result is a much larger contribution to the total polarizability from $\alpha_2$ compared to $\alpha_1$ when $|\gamma|$ is less than three and approaching equal contributions when $|\gamma|$ is larger than three (figure 1). $\alpha_1$ will always stay less than $\alpha_2$ even with very large $|\gamma|$ since $\alpha_2$ is associated with the larger radius, r.

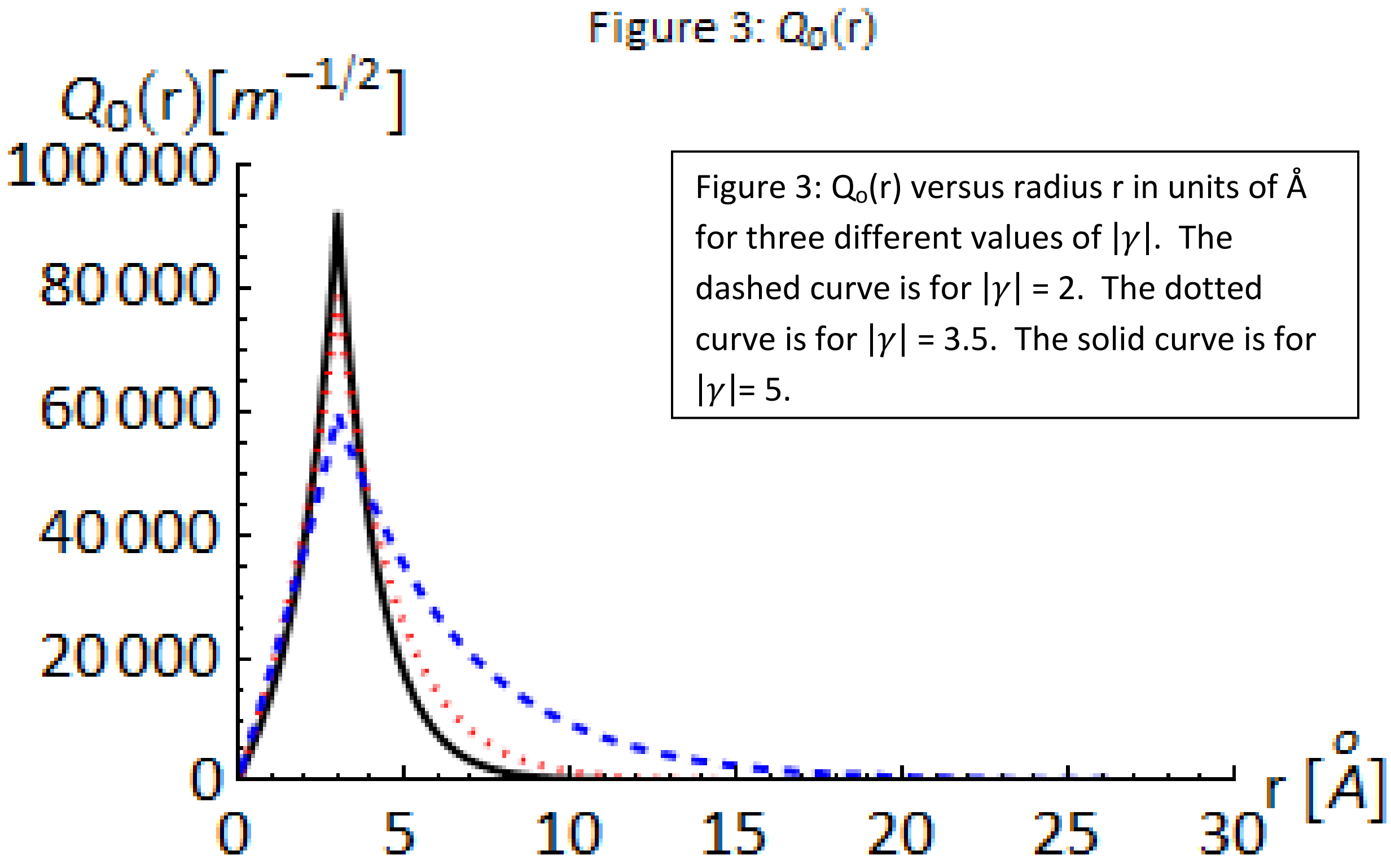

Figure 3: $Q_o(r)$ versus radius r in units of Å for three different values of $|\gamma|$. The dashed curve is for $|\gamma|$ = 2. The dotted curve is for $|\gamma|$ = 3.5. The solid curve is for $|\gamma|$= 5.

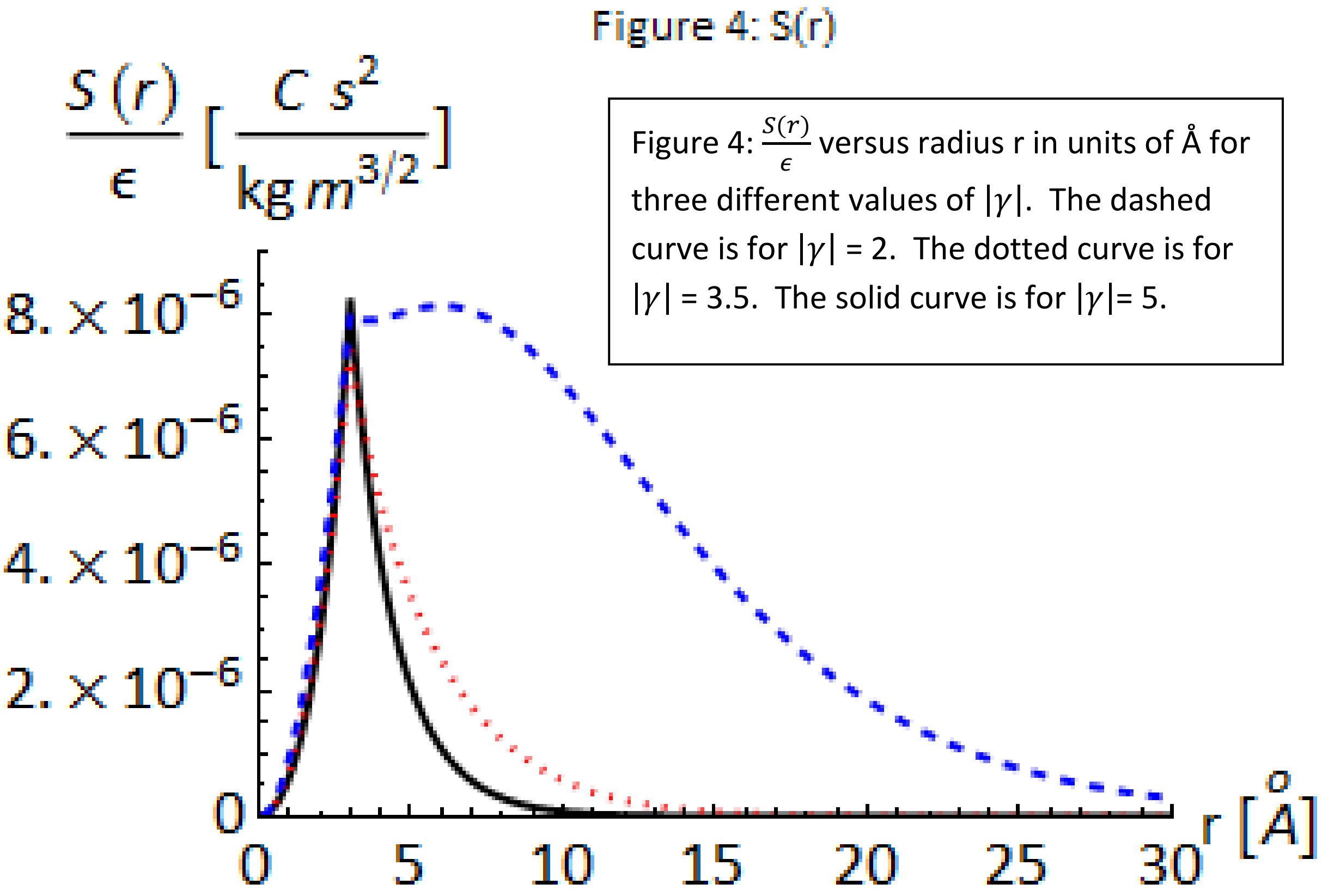

Figure 4: $\frac{S(r)}{\epsilon}$ versus radius r in units of Å for three different values of $|\gamma|$. The dashed curve is for $|\gamma|$ = 2. The dotted curve is for $|\gamma|$ = 3.5. The solid curve is for $|\gamma|$= 5.

## V. CONCLUSION

In this paper we employed the model of the spherical delta potential to calculate the energy shift of the ground state in second order perturbation theory. The energy shift which is the result of the interaction between a charged particle occupying the ground state and an applied static electric field was then used to find the electric polarizability of the bound system. In performing our perturbation calculations, we applied the Dalgarno-Lewis method for which we need only the unperturbed ground state rather than applying the conventional method which demands an infinite sum over all bound and continuum states. As illustrated in sections II and III, we did not need more than the knowledge of our undergraduate courses to find the ground state and determine the solution for the inhomogeneous differential equation. The elegance of the Dalgarno-Lewis method is in its simplicity which allows us to use our basic education of quantum mechanics in advancing our understanding of perturbation theory.

As in the one-dimensional case[1-3], we learn valuable lessons in studying the three-dimensional problem. In avoiding the full solution of the wave equation (Eq.(7)), we concentrate on finding the differential equation for the ground state (Eq.(8)). Then in solving Eq.(8) for $Q_o(r)$, we used our experience of solving the one-dimensional problem. The solution of the homogeneous version of Eq.(14) produces $Q_1(r)$. $Q_o(r)$ is used to find $\alpha$ while $Q_1(r)$ with $Q_o(r)$ are applied to Eq.(1) to find $\alpha_b$. $\alpha_b$ is then used to check our results for $\alpha$. In essence the Dalgarno-Lewis method can produce more than the ground state and gives us the necessary states which are needed for our check.

In our previous work[3] in which we studied the polarizability in the model of the finite potential well, we succeeded in separating the contributions to the polarizability from the classically forbidden and the classically allowed regions. The Dalgarno-Lewis method which allowed us to do that, allowed us in this work to separate the contribution to the electric polarizability from the region of $r < r_0$ and the region of $r > r_0$. The separation in the two problems would be impossible if we use the conventional method. At the same time studying the behavior of the system for small and large r which we did in this work is basic in many of Physics problems.

In conclusion the method of the inhomogeneous differential equation deserves our attention at the undergraduate and graduate levels. As our experience tells us it can be a great source for undergraduate research.

## VI. REFERENCES

a) Email: anis.maize@stvincent.edu
b) Email: john.smetanka@stvincent.edu

## APPENDIX A: COEFFICIENTS (C AND D) AND EXPRESSION FOR TOTAL AND BOUND TO BOUND POLARIZABILITY ($\alpha$ AND $\alpha_{BB}$)

$$C = -q\,\frac{N_0}{\sqrt{4\pi}}\left(\frac{2m}{\hbar^2}\right)\frac{(k_0^3 r_0^3 (3 + 2k_0 r_0 (3 + k_0 r_0)) + 3\,(1 + k_0 r_0)\gamma)\,Sinh(k_0 r_0) - (k_0 r_0\,(-3\gamma + k_0 r_0\,(3 + 2\gamma + 2k_0 r_0 (3 + k_0 r_0 + \gamma))))\,Cosh(k_0 r_0)}{4\,k_0^3\,((1 + k_0 r_0)^2 \gamma + e^{2k_0 r_0}(-\gamma + (k_0^2 r_0^2 (2k_0 r_0 + \gamma)))}$$

$$D = -q\,\frac{N_0}{\sqrt{4\pi}}\left(\frac{2m}{\hbar^2}\right)\frac{\begin{pmatrix} -12k_0^4 r_0^4 (Cosh(k_0 r_0))^2 + 2\,(2k_0^5 r_0^5 + k_0^3 r_0^3 (3 - 2\gamma) + 3\gamma + 3k_0 r_0 \gamma)(Sinh(k_0 r_0))^2 + \\ k_0 r_0\,(-3\gamma + k_0 r_0\,(-3\gamma + k_0 r_0 (3 + 2k_0 r_0 (-3 + k_0 r_0 + \gamma))))\,Sinh(2k_0 r_0) \end{pmatrix}}{16\,k_0^3 (k_0 r_0 (\gamma + k_0 r_0 (k_0 r_0 + \gamma))\,Cosh(k_0 r_0) + (k_0^3 r_0^3 - (1 + k_0 r_0)\gamma)\,Sinh(k_0 r_0))}$$

$$\alpha = \frac{mq^2}{3\hbar^2 \pi}\,\frac{\begin{pmatrix} -3 + k_0 r_0\,(-12 + k_0 r_0\,(-21 + 2k_0 r_0\,(-5 + 4k_0 r_0 (1 + k_0 r_0))))\,\gamma \\ +2e^{2k_0 r_0}\,(k_0^3 r_0^3\,(15 + 2k_0 r_0 (15 + 4k_0 r_0 (3 + k_0 r_0))) + (3 + k_0 r_0\,(6 + k_0 r_0\,(6 + k_0 r_0 (-1 + 2k_0 r_0 (1 + 2k_0 r_0)))))\,\gamma) \\ +e^{4k_0 r_0}\,(-3\gamma + k_0^2 r_0^2\,(-3\gamma + 2k_0 r_0 (-15 + 2k_0 r_0 (6k_0 r_0 + \gamma)))) \end{pmatrix}}{16k_0^4 (1 - e^{2k_0 r_0} + 2k_0 r_0)(1 + k_0 r_0)^2 \gamma + e^{2k_0 r_0}(-\gamma + k_0^2 r_0^2 (2k_0 r_0 + \gamma))}$$

$$\alpha_b = \frac{64mq^2}{3\hbar^2}\,\frac{\left(-e^{-2(k_0+k_1)r_0} k_0 \begin{pmatrix} -e^{2k_1 r_0} k_1^3 (-2 - 2k_0 r_0 - k_0^2 r_0^2 + k_1^2 r_0^2) + e^{2(k_0+k_1)r_0} (k_0 - k_1)^2 (-k_0 - 2k_1 + k_0 k_1^2 r_0^2 + k_1^3 r_0^2) + \\ e^{2k_0 r_0} k_0 (1 + k_1 r_0)(k_0^2 (1 + k_1 r_0) - k_1^2 (3 + k_1 r_0)) \end{pmatrix}^2\right)}{(k_0 - k_1)^5 k_1 (k_0 + k_1)^5 (1 - e^{2k_0 r_0} + 2k_0 r_0)(3 + 6k_1 r_0 + 5k_1^2 r_0^2 + 2k_1^3 r_0^3 + e^{2k_1 r_0}(-3 + k_1^2 r_0^2))}$$